\documentclass[10pt,aps,physrev,twocolumn,groupedaddress]{revtex4-2}

\usepackage{xcolor}
\usepackage{amsmath}
\usepackage{graphicx}
\usepackage{amssymb}
\usepackage{braket}
\usepackage{placeins}
\usepackage{appendix}
\usepackage{hyperref}
\hypersetup{
    colorlinks=true,
    linkcolor=blue,
    citecolor=blue,
    urlcolor=blue,
    pdftitle={Fractal Topology of Majorana Bound States in Superconducting Quasicrystals},
}

\begin{document}

\title{Fractal Topology of Majorana Bound States in Superconducting Quasicrystals}

\author{William Caiger}
\affiliation{School of Physics, Tyndall Avenue, Bristol, BS8 1TL, United Kingdom}
\author{Felix Flicker}
\affiliation{School of Physics, Tyndall Avenue, Bristol, BS8 1TL, United Kingdom}
\author{Miguel-\'Angel S\'anchez-Mart\'inez}
\email{contact@masanchezmartinez.com}
\affiliation{School of Physics, Tyndall Avenue, Bristol, BS8 1TL, United Kingdom}

\date{\today}

\begin{abstract}
Quasicrystalline order induces a fractal energy spectrum, yet its impact on topological protection remains an open fundamental question. Here, we demonstrate that the topological phase transitions characterised by the appearance of Majorana Bound States themselves have a fractal character. By extending this analysis to the full family of Sturmian words, we uncover \emph{Kitaev’s Butterfly} --- a spectral fractal analogous to Hofstadter’s butterfly, but fundamentally distinguished by a central superconducting gap. Within this framework, we identify \emph{Majorana’s Butterfly} as a fractal topological phase diagram governed by the competition between quasicrystallinity and superconducting pairing. We show that this competition dictates a hierarchy of Majorana stability, where the survival of the topological phase against fractal fragmentation is determined by the relative strength of these competing energy scales.
\end{abstract}

\maketitle

\section{\label{Introduction}Introduction}
Majorana Bound States (MBS) localised at the ends of 1D topological superconductors provide a possible route to fault-tolerant quantum computing~\cite{kitaev_unpaired_2001, leijnse_introduction_2012,sarma_majorana_2015,sato_topological_2017, lutchyn_majorana_2018, yazdani_hunting_2023}. In the periodic Kitaev chain~\cite{kitaev_unpaired_2001}, the topological phase hosting MBS occupies a single connected interval of chemical potential and terminates with a topological phase transition characterised by a gap closing in which bulk states reach zero energy. However, the introduction of quasiperiodicity into a bulk spectrum generically fragments it into a Cantor-set-like hierarchy of gaps~\cite{bellissard_spectral_1989, Flicker15, flicker_quasiperiodicity_2015, mace_critical_2017, kobialka_topological_2024}, raising fundamental questions regarding how this spectral fractality impacts the stability and transitions of the topological phase hosting MBS. Previous works have established the self-similarity of the order parameter as a function of the relative strength of quasicrystal parameters~\cite{ghadimi_majorana_2017}, but a more general understanding of the fractality of the topological phases and the underlying mechanisms is missing.

In this work we show that in Quasicrystal Kitaev Chains (QKCs) the sequence of transitions between topological and trivial phases, as a function of chemical potential, is itself fractal. This phenomenon arises from a direct competition between quasicrystallinity (QC) and superconductivity (SC). Using the Majorana polarisation as a real-space topological indicator, we show that the fractal gap structure of the bulk energy spectrum manifests in the chemical potential dependence, creating a hierarchy of gaps in the topological phase diagram. We capture this competition through a simple figure of merit: the difference between the quasicrystalline ($\Delta E_{QC}$) and superconducting ($\Delta E_{SC}$) gap sizes. 

Our results establish that only those quasicrystalline gaps satisfying $\Delta E_{QC}>\Delta E_{SC}$ survive the projection to zero energy and break the topological phase hosting MBS. Smaller QC gaps do not destroy the topological phase, but induce a hierarchy of finite hybridisations of the MBS. We generalise QKCs to a family of two-hopping models generated by \emph{Sturmian words}~\cite{morse_symbolic_1940, lothaire_sturmian_2002}~(Fig.~\ref{fig:Kitaev_model}). This construction reveals a similarity to the fractal $\mathbb{Z}$-invariant topology of Hofstadter's butterfly, albeit with a topologically nontrivial superconducting gap in the central region. We characterise the collection of bulk gap spectra as \emph{Kitaev’s butterfly}, and the resulting fractal topological phase diagram containing MBS as \emph{Majorana’s butterfly} --- a tunable fractal subset controlled by the QC-SC competition ratio.

\section{\label{sec:Model}The Model}

\begin{figure}[t]
\centering
    \includegraphics[width=\linewidth]{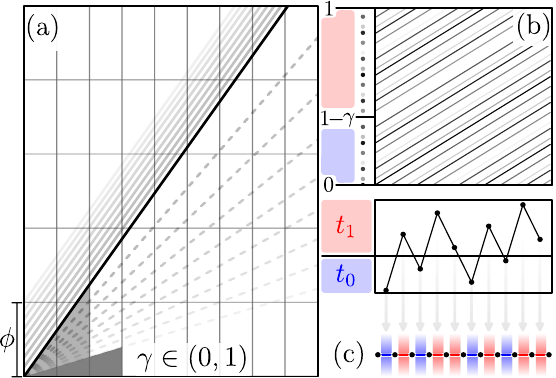}
    \caption{\textbf{(a)} A line of irrational slope $\gamma\in(0,1)$ and real $y$-intercept $\phi\in[0,1)$, modulo the periodic 2D integer lattice, defines a flow over a two-torus \textbf{(b)}. We project this into two intervals $(0;1-\gamma)\to t_0$ (blue) and $(1-\gamma;1)\to t_1$ (red) and use $t_0$, $t_1$ as hoppings in the Kitaev chain, shown in \textbf{(c)} for $L=11$. Each $\gamma$ defines a local isomorphism class of Sturmian words distinguished by $\phi$.}
    \label{fig:Kitaev_model}
\end{figure}

We generate 1D quasicrystals from the projection of an irrational slope in 2D Euclidean space~\cite{morse_symbolic_1940}. Of the various equivalent constructions~\cite{lothaire_sturmian_2002, bedaride_classification_2007},  Fig.~\ref{fig:Kitaev_model} shows how a straight line of slope $\gamma$ and vertical offset $\phi$ defines a linear flow over the two-torus when taken modulo the integer lattice. After projecting this flow onto a discrete set of points (Fig.~\ref{fig:Kitaev_model}(b)), each point is categorised based on intervals $(0;1-\gamma)$ and $(1-\gamma;1)$ \cite{bedaride_classification_2007} (Fig.~\ref{fig:Kitaev_model}(c)). The resulting sequence is long-range ordered, owing to the 2D periodic lattice, and is aperiodic since $\gamma$ is irrational~\cite{baake_aperiodic_2013}; we call these 1D quasicrystals (Refs.~\onlinecite{flicker_time_2018,Zaporski19,BoyleSteinhardt22} characterise highly symmetric examples, but we consider all cases here). Throughout this work we necessarily use finite approximants to true quasicrystals. To simulate quasiperiodicity in finite tilings of length $L$, we let $\gamma = a/b$ ($a$, $b$ integer) be a rational approximant to a true irrational. The word (sequence of $t_0$, $t_1$) generated by $\gamma$ has periodicity $b$, and so $b\geq L$ ensures the word appears quasiperiodic in the finite system~\cite{lothaire_sturmian_2002, baake_aperiodic_2013, bellissard_gap_1992}. Taking sufficiently long approximants allows us to represent physically relevant systems, and concomitantly to infer the behaviour of true quasicrystals in the thermodynamic limit from our results. This is the relevant sense of quasicrystal referred to hereafter. Fixing $\gamma$ and varying the $y$-intercept $\phi$ generates the `local isomorphism class' of the same quasicrystal: two quasicrystals are locally isomorphic if every finite sequence appearing in one appears in the other~\cite{baake_aperiodic_2013,Singh24}. Varying $\phi$ leads to structural re-arrangements called phasons~\cite{baake_aperiodic_2013}.

The Kitaev Hamiltonian for spinless electrons is given by~\cite{kitaev_unpaired_2001} 
\begin{equation}\label{eq:general_H}
    H = \sum_{j=1}^{L} - t_j (c_{j+1}^{\dag}c_j^{\phantom{\dag}} + \text{H.c.}) - \mu c_j^{\dag} c_j^{\phantom{\dag}} + (\Delta e^{i\theta} c_j c_{j+1} + \text{H.c.}),
\end{equation}
where $L$ is the number of sites, $j$ is the site index, the hopping potential is $t_j$, the on-site potential is $\mu$, and the superconducting pairing is $\Delta e^{i\theta}$. For the periodic model $t_j = t$, a 1D reciprocal space permits the straightforward definition of a $\mathbb{Z}_2$ momentum-space topological invariant that characterises the trivial and topological phases~\cite{kitaev_unpaired_2001}. This establishes a phase diagram for $\Delta, t \neq 0$ over $|\mu|<2t$ hosting MBS, with topological phase transitions at $|\mu_c|=2t$~\cite{kitaev_unpaired_2001}. In the following we will refer to the topologically non-trivial phases hosting MBS as `MBS phases'. The bandwidth of the periodic model is $W=4t$, and $\mu_c=2t$ is the critical on-site potential at which the bulk states occupy zero energy, breaking the edge localisation protecting MBS.

To construct a QKC, we take $\mu$ and $\Delta e^{i\theta}$ to be fixed along the chain (we take $\theta=0$ without loss of generality~\footnote{When $\theta=0,\pi$, the periodic Kitaev chain has time reversal symmetry and belongs to class BDI with a $\mathbb{Z}$ topological invariant, suggesting the possibility of 2 or more orthogonal MBS at each end. It has been shown that in order to have more than one unpaired Majorana mode at the end of a single-channel 1D superconductor like that described by Eq.~\eqref{eq:general_H}, modes emerge when adding long-range hoppings~\cite{degottardi2013majorana}, which we do not consider here. Additionally, the realistic models featuring spinful electrons, of which Eq.~\eqref{eq:general_H} is a simplification describing one gapped-out spin species, do not feature true Time reversal Symmetry and do not belong to class BDI~\cite{tewari_topological_2012}. For other values of $\theta$, the isolated chain presented in Eq.~\eqref{eq:general_H} is in class D, with a $\mathbb{Z}_2$ topological invariant corresponding to the presence or absence of MBS, and $\theta$ can be absorbed in the definition of the Majorana operators~\cite{kitaev_unpaired_2001}. }~\cite{kitaev_unpaired_2001, tewari_topological_2012,degottardi_majorana_2013,kobialka_topological_2024}), and $t_j$ to vary according to the construction in Fig.~\ref{fig:Kitaev_model}. Specifically, 
\begin{equation}\label{eq:cut-and-project_element}
   t_j = t_0 + (t_1 - t_0)\left(\lfloor(j+1)\gamma+\phi\rfloor - \lfloor j\gamma+\phi\rfloor\right),
\end{equation}
where $\gamma \in (0,1)$, $\phi\in[0,1)$. All $t_j$ are either $t_0$ or $t_1$. By normalising the physical parameters in Eq.~\eqref{eq:general_H} by $t_0$, we define dimensionless parameters $\mu' = \mu/t_0$, $\Delta' = \Delta/t_0$ and $\rho = t_1/t_0$. We observe that in the finite-length QKC, the bandwidth $W_{\gamma}\approx4\bar{t}_\gamma$, where $\bar{t}_\gamma=(1-\gamma) + \gamma \rho$, leading to a new critical chemical potential $|\mu'_c| \approx 2\bar{t}_\gamma$ consistent with the modified bandwidth. 

The bulk energy spectrum of the periodic Kitaev chain is continuous in the limit $L\to\infty$, whereas energy spectra of quasicrystals have a fractal structure featuring persistent energy gaps. This has been shown in the quasiperiodic Aubry-Andr\'e (AAH) model based on the Fibonacci QC~\cite{flicker_quasiperiodicity_2015,mace_fractal_2016, mace_critical_2017, jagannathan_fibonacci_2021}, which is equivalent to the QKC model in Eq.~\eqref{eq:general_H}, with $\gamma=\varphi^{-1}$ (where $\varphi$ is the golden ratio), in the limit $\Delta \to 0$. The manifestation of the QC spectral fractality in the topological phase, however, is not trivial, since not all QC gaps result in MBS-phase gaps. These quasicrystal energy gaps contain isolated mid-gap states which wind across the gap as $\phi$ is varied~\cite{kobialka_topological_2024} (detailed in Appendix~\ref{app:phasonWinding}). Through a full period of $\phi$ they wind an integer number of times, classifying each gap according to such a winding number~\cite{jagannathan_fibonacci_2021, kraus_topological_2012, verbin_observation_2013, huang_quantum_2018, huang_comparison_2019}. This can be efficiently calculated via the integrated density of states $N(E)$ using the gap-labelling theorem~\cite{bellissard_spectral_1989, bellissard_gap_1992,flicker_quasiperiodicity_2015,mace_gap_2017,jagannathan_fibonacci_2021}
\begin{equation}\label{eq:energy_gap_labelling}
    N(E) = p + \gamma q,
\end{equation}
where integers $(p,q)$ define a unique label for each energy gap, and $q$ is the mid-gap state winding number giving a $\mathbb{Z}$ invariant~\footnote{For continuously varying potentials this invariant is derived from the Chern number in the parent 2D lattice~\cite{flicker_quasiperiodicity_2015}, solving the $q$-sign ambiguity faced when using the Diophantine equation, Eq.~\eqref{eq:energy_gap_labelling}, alone~\cite{avron2014study}. However, for discontinuously varying potentials, as we have in Eq.~\eqref{eq:general_H}, the spectral projection is non-differentiable (see Appendix~\ref{app:phasonWinding}) and so the Berry phase is not well-defined~\cite{band2025quasiperiodicity}, invalidating any reference to the sign of the Berry phase to overcome the $q$-sign ambiguity. For our purposes here, we do not encounter $q$-sign ambiguity at the numerical resolution used, so we proceed with the gap-labelling theorem alone.}. In this labelling scheme the central superconducting gap receives the trivial $\mathbb{Z}$ label $q=0$ since the MBS mid-gap states do not exhibit winding behaviour. However, this gap exhibits a distinct $\mathbb{Z}_2$ topology identifying the existence of MBS.

\section{\label{sec:SingleKQCFractality}QC-SC Competition in a single Quasicrystal Kitaev Chain}

\begin{figure}
    \centering
    \includegraphics[width=\linewidth]{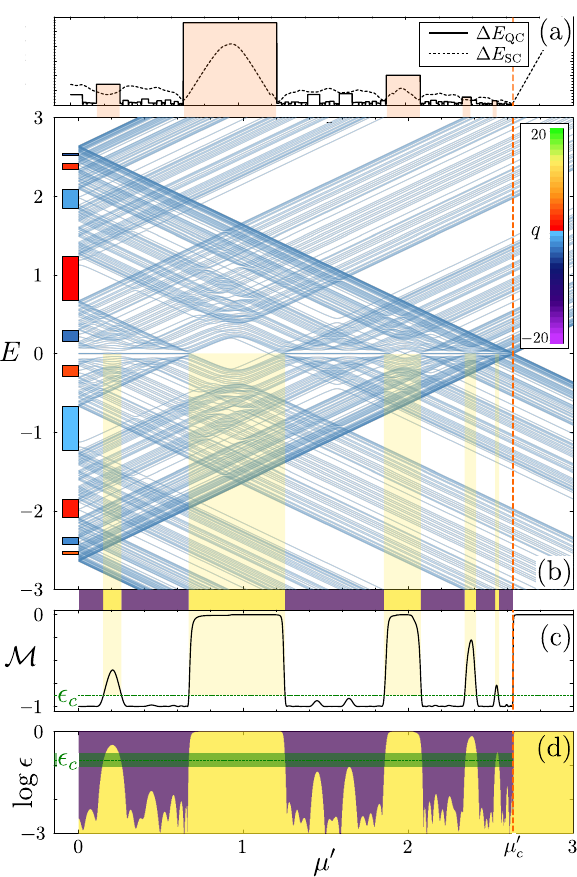}
    \caption{Results for the Fibonacci QKC $\gamma = \varphi^{-1}$ with $L=200$, $\rho=1.5$, $\Delta=0.05$, over $0 \leq \mu' \leq 3$ including $\mu'_c\approx2.61$. \textbf{(b)} shows the full fractal energy spectrum. Along $\mu'=0$ the QC energy gaps satisfying the QC-SC condition from Eq.~\eqref{eq:QC-SCCompCriterion} are coloured by the topological winding number $q$ from Eq.~\eqref{eq:energy_gap_labelling}. \textbf{(a)} shows QC wins this criterion in the five shaded regions; these gaps survive projection down to zero energy. They correspond to regions of $\mu'$ where $\mathcal{M}>-1+\epsilon_c$ defines trivial MBS phase, shown by the shaded (yellow) regions between \textbf{(b)} \& \textbf{(c)}. \textbf{(d)} shades topological regions (purple) and trivial regions (yellow) defined by $\mathcal{M}$ as $\log\epsilon$ is varied; at $\epsilon=\epsilon_c$ the expected five MBS phase gaps are realised, below this, further structure in the MP contains information about the hierarchy of finite MBS hybridisation induced by weaker QC gaps.}
    \label{fig:Figure 2}
\end{figure}

We first study how the strengths of quasicrystallinity (QC) and superconductivity (SC) compete in a single QKC for arbitrary $\gamma \in (0,1)$. We take $\gamma=\varphi^{-1}$ as a representative case whose corresponding electronic tight-binding model is known to have a multifractal density of states associated with the QC gaps~\cite{mace_fractal_2016, jagannathan_fibonacci_2021}. 

The MBS phase of the QKC cannot straightforwardly be characterised with reciprocal space methods owing to the lack of periodicity (although a 2D $k$-space description exists~\cite{baake_aperiodic_2013}). Moreover, aperiodicity can give rise to trivial zero-energy states~\cite{bagrets_class_2012, liu_andreev_2017}. Hence the direct detection of the topology of MBS cannot be achieved by spectral or state localisation analysis alone; despite the empirical correspondence between such signatures and MBS phases~\cite{akhmerov_quantized_2011, fulga_scattering_2012, hegde_majorana_2016}, experimental attempts demonstrate ambiguity when distinguishing MBS from other trivial zero-energy modes~\cite{deng_majorana_2016, vuik_reproducing_2019, pan_physical_2020}. Therefore, a real-space quantity that accounts for properties exclusive to MBS is needed. The Majorana Polarisation (MP) has been shown to effectively distinguish between MBS and trivial zero-energy states~\cite{sticlet_spin_2012, sedlmayr_visualizing_2015, awoga_identifying_2024}.

The MP tracks the particle-antiparticle equivalence of Majorana states between the left $(L)$ and right $(R)$ halves of the chain through the particle-hole operator $\hat{C}$. The MP is defined as $\mathcal{M} = P_L \cdot P_R^* = P_L^* \cdot P_R$, where
\begin{equation}\label{eq:MP_definition}
    \begin{split}
        P_L &= \frac{\sum_{j=1}^{L/2} \bra{\psi_{j,n}} \hat{C} \ket{\psi_{j,n}}}{\sum_{j=1}^{L/2} \braket{\psi_{j,n} | \psi_{j,n}}}, \\
        P_R &= \frac{\sum_{j=(L/2)+1}^{L} \bra{\psi_{j,M-n}} \hat{C} \ket{\psi_{j,M-n}}}{\sum_{j=(L/2)+1}^{L} \braket{\psi_{j,M-n} | \psi_{j,M-n}}},
    \end{split}
\end{equation}
and $M=2L+1$. The particle-hole operator $\hat{C}= e^{i\zeta}\hat{\tau}_x\hat{\mathcal{K}}$ is constructed for an arbitrary phase $\zeta$;  $\hat{\tau}_x$ is the Pauli-$x$ operator, and $\hat{\mathcal{K}}$ is the complex conjugate operator. Assuming energy eigenstates can be well-ordered by their corresponding eigenvalues, calculating the local action of $\hat{C}$ for $n=L$ compares the two candidate MBS eigenstates $\ket{\psi_{j,n}}$ with lowest absolute energy.

A value of $\mathcal{M} = -1$ indicates non-overlapping MBS localised at either end of the chain, with zero energy \cite{kobialka_topological_2024}. Notably, as shown in Fig.~\ref{fig:Figure 2}, $\mathcal{M}(\mu)$ is a continuous function, and so a finite tolerance for the deviation of $\mathcal{M} < -1 + \epsilon$ must be defined to use $\mathcal{M}$ to distinguish between the trivial and non-trivial Majorana topological phases. Hence, the topological phase classification is sensitive to the choice of $\epsilon$. This can be seen clearly in Fig.~\ref{fig:Figure 2}~(d): as $\epsilon$ is decreased, more gaps in the MBS phase are detected. 

The competition between QC and SC in the QKC model is key to understanding how the MBS phase transition itself inherits the fractal gap structure of the energy spectrum. Fig.~\ref{fig:Figure 2} shows how the QC energy gaps at $\mu'=0$ are projected onto the $\mu'$ axis, causing gaps in the MBS phase. This projection is not trivial: topologically trivial regions only occur when the QC energy gaps overpower the SC gap. We refer to these as gaps in the MBS phase. Smaller QC energy gaps do not survive to zero energy. Instead, they are \emph{reflected} by the SC gap. A particular QC energy gap with size $\Delta E_\text{QC}$ results in an MBS phase gap if it is larger than the SC gap formed by the bulk states $\Delta E_\text{SC}$ in the $\mu'$ region onto which it is projected. Fig.~\ref{fig:Figure 2}~(a) shows both gap sizes $\Delta E_{QC}$ and $\Delta E_{SC}$ along $\mu'$, shading regions where
\begin{equation}\label{eq:QC-SCCompCriterion}
    \Delta E_\text{QC} > \Delta E_\text{SC}. 
\end{equation}
For these regions, the QC gap strength overpowers the SC gap strength and the MBS phase is broken. Finally, this competition criterion defines an expected number of gaps in the MBS phase which can be used to obtain an appropriate value of $\epsilon_c$, shown in Fig.~\ref{fig:Figure 2}~(d). 

Therefore, of the fractal energy spectrum with finite scale controlled by the system size $L$, a subset of gaps (Fig.~\ref{fig:Figure 2}~(b)) can be taken according to the QC-SC competition criterion $\Delta E_\text{QC} > \Delta E_\text{SC}$. This subset defines the gaps in the MBS phase, with topological phase transitions at the edges of such gaps, resulting in a finite-scale fractal topological phase diagram controlled by $\rho/\Delta'$. That is, varying $\rho/\Delta'$ tunes the relative strength of QC to SC, adjusting the number of gaps that fulfil the competition criterion. The true fractal is achieved in the limit $\rho/\Delta' \to \infty$ in which all QC energy gaps are projected. 

It is important to note that for small chain lengths finite-size effects might become relevant, and a careful analysis of such effects is needed to establish a reliable value of $\epsilon_c$ charaterising trivial and topological phases. Notably, for those gaps which fulfil the competition criterion, the isolated mid-gap state of the QC energy gap is also projected into the MBS phase gap at zero energy. Importantly, these projected zero-energy mid-gap states do not cause the value of MP to drop back towards $\mathcal{M}<-1+\epsilon$, confirming they are not MBS. Rather, they are trivial zero-energy states. This demonstrates how quasicrystallinity can be a source of trivial zero-energy states in a QKC. For more details see Appendix~\ref{app:phasonWinding}.

Finally, Fig.~\ref{fig:Figure 2}~(d) reveals that the MP contains more structure than is apparent from only looking at the MP discretised by the tolerance $\epsilon_c$. These regions of $\mathcal{M}<-1+\epsilon$ below the appropriate tolerance range contain information about the hybridisation of MBS. In finite systems, there is always a non-zero hybridisation between MBS, hence, even when $\Delta E_\text{QC} \not> \Delta E_\text{SC}$ the QC gaps induce a degree of hybridisation in the MBS states quantified by the MP. The height hierarchy of these \emph{domes} indicates the order in which regions in $\mu'$ would become MBS phase gaps as $\rho/\Delta'$ is increased. In the limit $\rho/\Delta'\to\infty$ there is no height hierarchy between domes and all QC energy gaps satisfy $\Delta E_\text{QC} > \Delta E_\text{SC}$, being realised as MBS phase gaps. Furthermore, for sufficiently large approximants that avoid finite-size effects, the hierarchy set by $\rho/\Delta'$ is unchanged as $L$ is increased, implying this behaviour will be consistent in the true quasiperiodic limit. For more details see Appendix~\ref{app:hierarchy}.

\section{\label{sec:Generalising}Generalising to all Quasicrystal Kitaev ChainS}

\begin{figure*}[t]
    \centering
    \includegraphics[width=\textwidth]{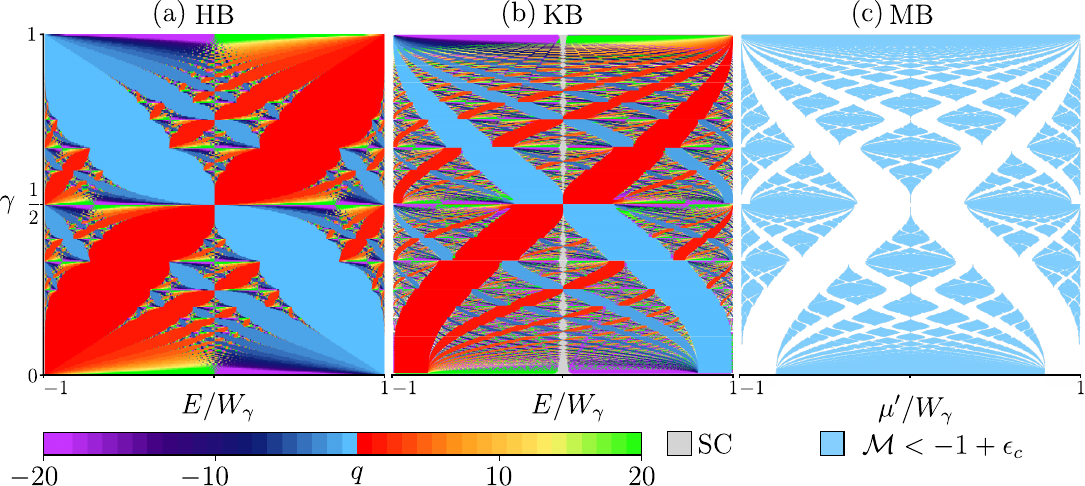}
    \caption{\textbf{(a)} Hofstadter's butterfly (HB) and \textbf{(b)} Kitaev's butterfly (KB) both normalised to their bandwidth $(W_\gamma)$, and coloured by $q$-label values, fixing $|q_\text{max}|=20$ for visual clarity. KB here is plotted for $L=500$, $\mu'=0$, $\rho=2.0$ and $\Delta'=0.05$ for clarity. Notably, HB and KB exhibit the same antisymmetric labelling structure and behaviour around rational values of $\gamma$, except for the superconducting gap (SC) along $E=0$ (gray) in KB which is not labelled by $q$. \textbf{(c)} Majorana's butterfly (MB) is composed of regions in the MBS topological phase using the Majorana polarisation $\mathcal{M}$ for the same parameters as KB and with $\mu'$ normalised to the same bandwidth as KB. The fractal structure of the MBS phase transition in MB is a subset of KB with finite-scale controlled by the QC-SC competition ratio $\rho/\Delta'$.}
    \label{fig:Figure 3}
\end{figure*}

The Fibonacci QKC, with $\gamma =\varphi^{-1}$, is just one member of the family of quasiperiodic models with irrational $\gamma$. It has previously been shown that fractal spectral properties are consistent across all irrational $\gamma \in (0,1)$ (defining the class of Sturmian words)~\cite{bellissard_spectral_1989, damanik_uniform_1999, NPytheasFogg}. Therefore, not only is each individual QKC spectrum a fractal resembling a Cantor set, but the collection of spectra of all 1D QKCs combine to form another fractal structure. We show this fractal, which we call Kitaev's butterfly (KB), in Fig.~\ref{fig:Figure 3}~(b). We have coloured QC energy gaps according to the topological winding number $q$ from Eq.~\eqref{eq:energy_gap_labelling}.

We observe a topological similarity between KB and the well-known Hofstadter's butterfly (HB), shown in Fig.~\ref{fig:Figure 3}~(a). HB-like patterns have been reported in superconducting settings~\cite{ghadimi_majorana_2017,degottardi2013majorana}, but their topological content and connection to HB have not been established. Here, we identify the shared topological structure between HB and KB and highlight the key differences between them. We consider a normalisation of HB to aid the visual comparison with KB. HB consists of a fractal topological phase diagram arising from a collection of instances of the Harper model (equivalent to the periodic AAH model)~\cite{hofstadter_energy_1976}. In each instance, an irrational ratio of magnetic flux $\gamma = \Phi/\Phi_0$ is applied per unit cell of a 2D periodic lattice. The resulting spectrum is fractal in the limit of all $\gamma \to \mathbb{R}$ truly irrational values~\cite{hofstadter_energy_1976, simon_almost_1982}, with each energy gap being defined by a quantised Hall conductance~\cite{thouless_quantized_1982, kohmoto_topological_1985}. This quantised Hall conductance, $q$, is a topologically protected quantity characterised by the Chern number~\cite{dana_quantised_1985, bellissard_k-theory_1986}. Fig.~\ref{fig:Figure 3}~(a) shows the resulting fractal butterfly pattern of this topological phase diagram in which energy gaps are coloured by $q$.

Comparing the topological features of finite simulations of HB and KB, we observe the following. Firstly, $q$-labels in both HB and KB are antisymmetric about $\gamma=1/2$ and about $E=0$ such that $q(\gamma, E) = -q(1-\gamma, -E)$. Secondly, both HB and KB exhibit the same $q$-labelling behaviour around rational values $\gamma = a/n$ ($a$, $n$ integer, $n<L$) as the Farey sequence~\cite{abanov1998hierarchical, satija2016tale, beckus2025spectral}. Around these points the gap structure becomes dense (continuous in the limit $L\to\infty$) and labels reach $|q_\text{max}|$ (the maximum label permitted in the limited-resolution simulation for visual clarity). Such $\gamma$ have exactly $n$ continuous energy bands whose widths are ordered by $n$ and whose centres are given by $n^\text{th}$ order Chambers polynomials \cite{hofstadter_energy_1976, jitomirskaya2022spectrum}. The largest continuous bands are for $n=1$ at $\gamma=1/1$ and its complement $0/1$ centred at $E=0$, the next largest is for $n=2$ at $\gamma=1/2$ centred at $E=\pm 2$, and so on. Thirdly, KB features a $q=0$ SC gap around $E=0$ which is not present in HB. This SC gap hosts the MBS, a topological phase not present in HB. Hence, KB is a new fractal set which shares important topological structures with HB, but is not completely equivalent to it.

There are other non-topological differences between the shape of HB and KB as a result of the discontinuity of the varying potential in the QKC model. These have been understood in terms of a discontinuously varying chemical potential, giving rise to Kohmoto's butterfly~\cite{kohmoto1983localization, band2025quasiperiodicity}, a non-superconducting analogue of KB. Details of these differences are shown in Appendix~\ref{app:unnormalisedButterflies}.

It is important to clarify here that the apparently continuous limit of the structure when $q\to |q_\text{max}|$ is a product of finite numerical simulation size. In HB the precision of approximation of $\gamma$ to a real irrational value determines the finite numerical scale of the fractal pattern; but in KB the underlying QKC model is not periodic, so both the precision of $\gamma$ and the chain length $L$ determine the finite fractal scale. In the limit of $L \to \infty$ (assuming also appropriate numerical $\gamma$ precision), KB is expected to converge to a truly fractal structure. Crucially, for KB the resolution of fractal structure is not only limited by numerical precision, but also the physical parameter $L$. 

Given the projective relationship between the bulk energy structure of the QKC and the topological phase in the $\mu'$ axis established in the previous section, we naturally expect to observe some of the structure of KB in MBS phase diagram. We construct what we term Majorana's butterfly (MB), Fig.~\ref{fig:Figure 3}~(c), by sampling the $\mu'$ axis and categorising the topology at each point using the MP. MB shares structural features with KB, but with an important difference: even for $L\to\infty$ when KB is truly fractal, only a subset of the gaps in KB are projected onto MB --- specifically, only those that satisfy the QC-SC competition criterion. Thus MB has a different finite scale control given by the physical parameter ratio $\rho/\Delta'$. This means for finite $\rho/\Delta'$ presented here, the dense regions of MB around rational values $\gamma=a/n$, really are continuous regions of MBS phase. Just like in the individual QKC case, as $\rho/\Delta'$ is increased, finer fractal structure from KB is realised in MB with predictable order based on the hybridisation hierarchy. Therefore, MB constitutes a fractal subset containing the non-trivial competition between QC and SC, and can be seen as a non-trivial subset of KB.

\section{\label{sec:Conclusions}Conclusions}

We have shown that quasicrystalline order can render the topological phase structure fractal in 1D superconductors, in a non-trivial manner. Focusing on two-hopping Quasicrystal Kitaev Chains, and using the Majorana polarization as a real-space topological indicator, we demonstrate that the chemical potential phase diagram inherits a fractal hierarchy of gaps from the quasicrystalline bulk spectrum. This fractal structure of the topological phase diagram is not trivially inherited from the quasicrystal: it is controlled by a direct competition between quasicrystallinity and superconductivity.

One of our central results is a simple energetic criterion that predicts which quasicrystalline gaps survive projection to zero energy and thereby interrupt the Majorana phase: a quasicrystalline gap of size $\Delta E_{\mathrm{QC}}$ opens a gap in the Majorana phase only when it exceeds the local superconducting gap scale $\Delta E_{\mathrm{SC}}$ in the corresponding chemical-potential window, $\Delta E_{\mathrm{QC}}>\Delta E_{\mathrm{SC}}$. Gaps that fail to meet this criterion do not destroy the topological phase, but instead induce a structured hierarchy of finite Majorana hybridisations captured by the Majorana polarisation. This hierarchy can be used to fix the tolerance $\epsilon$ needed to classify phases with the Majorana polarisation in finite systems, and dictates the order in which additional topological phase transitions appear as the QC–SC competition ratio is tuned.

Extending from a single quasicrystal to the full family of Sturmian words, we uncovered Kitaev’s butterfly: the topologically-labelled collection of fractal spectra of Sturmian Quasicrystal Kitaev Chains, closely analogous to Hofstadter’s butterfly yet fundamentally distinguished by a central superconducting gap (along with other non-topological distinctions visible in Fig.~\ref{fig:Figure 3}). Projecting this structure onto the chemical potential yields Majorana’s butterfly, a tunable fractal subset that delineates where Majorana bound states survive. Experimentally, we therefore expect the fractal topological phase transitions to be accessible by electrostatic gating: tuning the carrier density shifts the effective chemical potential and should allow direct mapping of the fractal sequence of topological transitions. More broadly, the topological fractality we identify provides a clear fingerprint to distinguish Majorana bound states from other zero-energy modes that can arise in quasicrystals --- \emph{e.g.} mid-gap states forced by varying the phason degree of freedom $\phi$~\cite{Madsen13,flicker_quasiperiodicity_2015} --- by correlating the zero-energy response with the fractal structure of the underlying quasicrystalline spectrum.

\section{Acknowledgements}

F.F. and M.-Á.S.-M. acknowledge financial support from EPSRC grant EP/X012239/1. W.C., F.F., and M.-Á.S.-M. acknowledge the use of the computational facilities of the Advanced Computing Research Centre, University of Bristol.

% \bibliography{refs}

\begin{thebibliography}{60}%
\makeatletter
\providecommand \@ifxundefined [1]{%
 \@ifx{#1\undefined}
}%
\providecommand \@ifnum [1]{%
 \ifnum #1\expandafter \@firstoftwo
 \else \expandafter \@secondoftwo
 \fi
}%
\providecommand \@ifx [1]{%
 \ifx #1\expandafter \@firstoftwo
 \else \expandafter \@secondoftwo
 \fi
}%
\providecommand \natexlab [1]{#1}%
\providecommand \enquote  [1]{``#1''}%
\providecommand \bibnamefont  [1]{#1}%
\providecommand \bibfnamefont [1]{#1}%
\providecommand \citenamefont [1]{#1}%
\providecommand \href@noop [0]{\@secondoftwo}%
\providecommand \href [0]{\begingroup \@sanitize@url \@href}%
\providecommand \@href[1]{\@@startlink{#1}\@@href}%
\providecommand \@@href[1]{\endgroup#1\@@endlink}%
\providecommand \@sanitize@url [0]{\catcode `\\12\catcode `\$12\catcode `\&12\catcode `\#12\catcode `\^12\catcode `\_12\catcode `\%12\relax}%
\providecommand \@@startlink[1]{}%
\providecommand \@@endlink[0]{}%
\providecommand \url  [0]{\begingroup\@sanitize@url \@url }%
\providecommand \@url [1]{\endgroup\@href {#1}{\urlprefix }}%
\providecommand \urlprefix  [0]{URL }%
\providecommand \Eprint [0]{\href }%
\providecommand \doibase [0]{https://doi.org/}%
\providecommand \selectlanguage [0]{\@gobble}%
\providecommand \bibinfo  [0]{\@secondoftwo}%
\providecommand \bibfield  [0]{\@secondoftwo}%
\providecommand \translation [1]{[#1]}%
\providecommand \BibitemOpen [0]{}%
\providecommand \bibitemStop [0]{}%
\providecommand \bibitemNoStop [0]{.\EOS\space}%
\providecommand \EOS [0]{\spacefactor3000\relax}%
\providecommand \BibitemShut  [1]{\csname bibitem#1\endcsname}%
\let\auto@bib@innerbib\@empty
%</preamble>
\bibitem [{\citenamefont {Kitaev}(2001)}]{kitaev_unpaired_2001}%
  \BibitemOpen
  \bibfield  {author} {\bibinfo {author} {\bibfnamefont {A.~Y.}\ \bibnamefont {Kitaev}},\ }\href {https://doi.org/10.1070/1063-7869/44/10S/S29} {\bibfield  {journal} {\bibinfo  {journal} {Physics-Uspekhi}\ }\textbf {\bibinfo {volume} {44}},\ \bibinfo {pages} {131} (\bibinfo {year} {2001})}\BibitemShut {NoStop}%
\bibitem [{\citenamefont {Leijnse}\ and\ \citenamefont {Flensberg}(2012)}]{leijnse_introduction_2012}%
  \BibitemOpen
  \bibfield  {author} {\bibinfo {author} {\bibfnamefont {M.}~\bibnamefont {Leijnse}}\ and\ \bibinfo {author} {\bibfnamefont {K.}~\bibnamefont {Flensberg}},\ }\href {https://doi.org/10.1088/0268-1242/27/12/124003} {\bibfield  {journal} {\bibinfo  {journal} {Semiconductor Science and Technology}\ }\textbf {\bibinfo {volume} {27}},\ \bibinfo {pages} {124003} (\bibinfo {year} {2012})}\BibitemShut {NoStop}%
\bibitem [{\citenamefont {Sarma}\ \emph {et~al.}(2015)\citenamefont {Sarma}, \citenamefont {Freedman},\ and\ \citenamefont {Nayak}}]{sarma_majorana_2015}%
  \BibitemOpen
  \bibfield  {author} {\bibinfo {author} {\bibfnamefont {S.~D.}\ \bibnamefont {Sarma}}, \bibinfo {author} {\bibfnamefont {M.}~\bibnamefont {Freedman}},\ and\ \bibinfo {author} {\bibfnamefont {C.}~\bibnamefont {Nayak}},\ }\href {https://doi.org/10.1038/npjqi.2015.1} {\bibfield  {journal} {\bibinfo  {journal} {npj Quantum Information}\ }\textbf {\bibinfo {volume} {1}},\ \bibinfo {pages} {15001} (\bibinfo {year} {2015})}\BibitemShut {NoStop}%
\bibitem [{\citenamefont {Sato}\ and\ \citenamefont {Ando}(2017)}]{sato_topological_2017}%
  \BibitemOpen
  \bibfield  {author} {\bibinfo {author} {\bibfnamefont {M.}~\bibnamefont {Sato}}\ and\ \bibinfo {author} {\bibfnamefont {Y.}~\bibnamefont {Ando}},\ }\href {https://doi.org/10.1088/1361-6633/aa6ac7} {\bibfield  {journal} {\bibinfo  {journal} {Reports on Progress in Physics}\ }\textbf {\bibinfo {volume} {80}},\ \bibinfo {pages} {076501} (\bibinfo {year} {2017})}\BibitemShut {NoStop}%
\bibitem [{\citenamefont {Lutchyn}\ \emph {et~al.}(2018)\citenamefont {Lutchyn}, \citenamefont {Bakkers}, \citenamefont {Kouwenhoven}, \citenamefont {Krogstrup}, \citenamefont {Marcus},\ and\ \citenamefont {Oreg}}]{lutchyn_majorana_2018}%
  \BibitemOpen
  \bibfield  {author} {\bibinfo {author} {\bibfnamefont {R.~M.}\ \bibnamefont {Lutchyn}}, \bibinfo {author} {\bibfnamefont {E.~P. A.~M.}\ \bibnamefont {Bakkers}}, \bibinfo {author} {\bibfnamefont {L.~P.}\ \bibnamefont {Kouwenhoven}}, \bibinfo {author} {\bibfnamefont {P.}~\bibnamefont {Krogstrup}}, \bibinfo {author} {\bibfnamefont {C.~M.}\ \bibnamefont {Marcus}},\ and\ \bibinfo {author} {\bibfnamefont {Y.}~\bibnamefont {Oreg}},\ }\href {https://doi.org/10.1038/s41578-018-0003-1} {\bibfield  {journal} {\bibinfo  {journal} {Nature Reviews Materials}\ }\textbf {\bibinfo {volume} {3}},\ \bibinfo {pages} {52} (\bibinfo {year} {2018})}\BibitemShut {NoStop}%
\bibitem [{\citenamefont {Yazdani}\ \emph {et~al.}(2023)\citenamefont {Yazdani}, \citenamefont {Von~Oppen}, \citenamefont {Halperin},\ and\ \citenamefont {Yacoby}}]{yazdani_hunting_2023}%
  \BibitemOpen
  \bibfield  {author} {\bibinfo {author} {\bibfnamefont {A.}~\bibnamefont {Yazdani}}, \bibinfo {author} {\bibfnamefont {F.}~\bibnamefont {Von~Oppen}}, \bibinfo {author} {\bibfnamefont {B.~I.}\ \bibnamefont {Halperin}},\ and\ \bibinfo {author} {\bibfnamefont {A.}~\bibnamefont {Yacoby}},\ }\href {https://doi.org/10.1126/science.ade0850} {\bibfield  {journal} {\bibinfo  {journal} {Science}\ }\textbf {\bibinfo {volume} {380}},\ \bibinfo {pages} {eade0850} (\bibinfo {year} {2023})}\BibitemShut {NoStop}%
\bibitem [{\citenamefont {Bellissard}\ \emph {et~al.}(1989)\citenamefont {Bellissard}, \citenamefont {Iochum}, \citenamefont {Scoppola},\ and\ \citenamefont {Testard}}]{bellissard_spectral_1989}%
  \BibitemOpen
  \bibfield  {author} {\bibinfo {author} {\bibfnamefont {J.}~\bibnamefont {Bellissard}}, \bibinfo {author} {\bibfnamefont {B.}~\bibnamefont {Iochum}}, \bibinfo {author} {\bibfnamefont {E.}~\bibnamefont {Scoppola}},\ and\ \bibinfo {author} {\bibfnamefont {D.}~\bibnamefont {Testard}},\ }\href {https://doi.org/10.1007/BF01218415} {\bibfield  {journal} {\bibinfo  {journal} {Communications in Mathematical Physics}\ }\textbf {\bibinfo {volume} {125}},\ \bibinfo {pages} {527} (\bibinfo {year} {1989})}\BibitemShut {NoStop}%
\bibitem [{\citenamefont {Flicker}\ and\ \citenamefont {van Wezel}(2015{\natexlab{a}})}]{Flicker15}%
  \BibitemOpen
  \bibfield  {author} {\bibinfo {author} {\bibfnamefont {F.}~\bibnamefont {Flicker}}\ and\ \bibinfo {author} {\bibfnamefont {J.}~\bibnamefont {van Wezel}},\ }\href@noop {} {\bibfield  {journal} {\bibinfo  {journal} {Physical Review Letters}\ }\textbf {\bibinfo {volume} {115}},\ \bibinfo {pages} {236401} (\bibinfo {year} {2015}{\natexlab{a}})}\BibitemShut {NoStop}%
\bibitem [{\citenamefont {Flicker}\ and\ \citenamefont {van Wezel}(2015{\natexlab{b}})}]{flicker_quasiperiodicity_2015}%
  \BibitemOpen
  \bibfield  {author} {\bibinfo {author} {\bibfnamefont {F.}~\bibnamefont {Flicker}}\ and\ \bibinfo {author} {\bibfnamefont {J.}~\bibnamefont {van Wezel}},\ }\href {https://doi.org/10.1209/0295-5075/111/37008} {\bibfield  {journal} {\bibinfo  {journal} {Europhysics Letters}\ }\textbf {\bibinfo {volume} {111}},\ \bibinfo {pages} {37008} (\bibinfo {year} {2015}{\natexlab{b}})}\BibitemShut {NoStop}%
\bibitem [{\citenamefont {Macé}\ \emph {et~al.}(2017{\natexlab{a}})\citenamefont {Macé}, \citenamefont {Jagannathan}, \citenamefont {Kalugin}, \citenamefont {Mosseri},\ and\ \citenamefont {Piéchon}}]{mace_critical_2017}%
  \BibitemOpen
  \bibfield  {author} {\bibinfo {author} {\bibfnamefont {N.}~\bibnamefont {Macé}}, \bibinfo {author} {\bibfnamefont {A.}~\bibnamefont {Jagannathan}}, \bibinfo {author} {\bibfnamefont {P.}~\bibnamefont {Kalugin}}, \bibinfo {author} {\bibfnamefont {R.}~\bibnamefont {Mosseri}},\ and\ \bibinfo {author} {\bibfnamefont {F.}~\bibnamefont {Piéchon}},\ }\href {https://doi.org/10.1103/PhysRevB.96.045138} {\bibfield  {journal} {\bibinfo  {journal} {Physical Review B}\ }\textbf {\bibinfo {volume} {96}},\ \bibinfo {pages} {045138} (\bibinfo {year} {2017}{\natexlab{a}})}\BibitemShut {NoStop}%
\bibitem [{\citenamefont {Kobiałka}\ \emph {et~al.}(2024)\citenamefont {Kobiałka}, \citenamefont {Awoga}, \citenamefont {Leijnse}, \citenamefont {Domański}, \citenamefont {Holmvall},\ and\ \citenamefont {Black-Schaffer}}]{kobialka_topological_2024}%
  \BibitemOpen
  \bibfield  {author} {\bibinfo {author} {\bibfnamefont {A.}~\bibnamefont {Kobiałka}}, \bibinfo {author} {\bibfnamefont {O.~A.}\ \bibnamefont {Awoga}}, \bibinfo {author} {\bibfnamefont {M.}~\bibnamefont {Leijnse}}, \bibinfo {author} {\bibfnamefont {T.}~\bibnamefont {Domański}}, \bibinfo {author} {\bibfnamefont {P.}~\bibnamefont {Holmvall}},\ and\ \bibinfo {author} {\bibfnamefont {A.~M.}\ \bibnamefont {Black-Schaffer}},\ }\href {https://doi.org/10.1103/PhysRevB.110.134508} {\bibfield  {journal} {\bibinfo  {journal} {Physical Review B}\ }\textbf {\bibinfo {volume} {110}},\ \bibinfo {pages} {134508} (\bibinfo {year} {2024})}\BibitemShut {NoStop}%
\bibitem [{\citenamefont {Ghadimi}\ \emph {et~al.}(2017)\citenamefont {Ghadimi}, \citenamefont {Sugimoto},\ and\ \citenamefont {Tohyama}}]{ghadimi_majorana_2017}%
  \BibitemOpen
  \bibfield  {author} {\bibinfo {author} {\bibfnamefont {R.}~\bibnamefont {Ghadimi}}, \bibinfo {author} {\bibfnamefont {T.}~\bibnamefont {Sugimoto}},\ and\ \bibinfo {author} {\bibfnamefont {T.}~\bibnamefont {Tohyama}},\ }\href {https://doi.org/10.7566/JPSJ.86.114707} {\bibfield  {journal} {\bibinfo  {journal} {Journal of the Physical Society of Japan}\ }\textbf {\bibinfo {volume} {86}},\ \bibinfo {pages} {114707} (\bibinfo {year} {2017})}\BibitemShut {NoStop}%
\bibitem [{\citenamefont {Morse}\ and\ \citenamefont {Hedlund}(1940)}]{morse_symbolic_1940}%
  \BibitemOpen
  \bibfield  {author} {\bibinfo {author} {\bibfnamefont {M.}~\bibnamefont {Morse}}\ and\ \bibinfo {author} {\bibfnamefont {G.~A.}\ \bibnamefont {Hedlund}},\ }\href {https://doi.org/10.2307/2371431} {\bibfield  {journal} {\bibinfo  {journal} {American Journal of Mathematics}\ }\textbf {\bibinfo {volume} {62}},\ \bibinfo {pages} {1} (\bibinfo {year} {1940})}\BibitemShut {NoStop}%
\bibitem [{\citenamefont {Lothaire}(2002)}]{lothaire_sturmian_2002}%
  \BibitemOpen
  \bibfield  {author} {\bibinfo {author} {\bibfnamefont {M.}~\bibnamefont {Lothaire}},\ }in\ \href {https://doi.org/10.1017/CBO9781107326019.003} {\emph {\bibinfo {booktitle} {Algebraic {Combinatorics} on {Words}}}}\ (\bibinfo  {publisher} {Cambridge University Press},\ \bibinfo {year} {2002})\ pp.\ \bibinfo {pages} {45--110}\BibitemShut {NoStop}%
\bibitem [{\citenamefont {Bedaride}(2007)}]{bedaride_classification_2007}%
  \BibitemOpen
  \bibfield  {author} {\bibinfo {author} {\bibfnamefont {N.}~\bibnamefont {Bedaride}},\ }\href {https://doi.org/10.1016/j.tcs.2007.05.037} {\bibfield  {journal} {\bibinfo  {journal} {Theoretical Computer Science}\ }\textbf {\bibinfo {volume} {385}},\ \bibinfo {pages} {214} (\bibinfo {year} {2007})}\BibitemShut {NoStop}%
\bibitem [{\citenamefont {Baake}\ and\ \citenamefont {Grimm}(2013)}]{baake_aperiodic_2013}%
  \BibitemOpen
  \bibfield  {author} {\bibinfo {author} {\bibfnamefont {M.}~\bibnamefont {Baake}}\ and\ \bibinfo {author} {\bibfnamefont {U.}~\bibnamefont {Grimm}},\ }\href@noop {} {\emph {\bibinfo {title} {Aperiodic {Order}}}}\ (\bibinfo  {publisher} {Cambridge University Press},\ \bibinfo {year} {2013})\BibitemShut {NoStop}%
\bibitem [{\citenamefont {Flicker}(2018)}]{flicker_time_2018}%
  \BibitemOpen
  \bibfield  {author} {\bibinfo {author} {\bibfnamefont {F.}~\bibnamefont {Flicker}},\ }\href {https://doi.org/10.21468/SciPostPhys.5.1.001} {\bibfield  {journal} {\bibinfo  {journal} {SciPost Physics}\ }\textbf {\bibinfo {volume} {5}},\ \bibinfo {pages} {001} (\bibinfo {year} {2018})}\BibitemShut {NoStop}%
\bibitem [{\citenamefont {Zaporski}\ and\ \citenamefont {Flicker}(2019)}]{Zaporski19}%
  \BibitemOpen
  \bibfield  {author} {\bibinfo {author} {\bibfnamefont {L.}~\bibnamefont {Zaporski}}\ and\ \bibinfo {author} {\bibfnamefont {F.}~\bibnamefont {Flicker}},\ }\href@noop {} {\bibfield  {journal} {\bibinfo  {journal} {SciPost Physics}\ }\textbf {\bibinfo {volume} {7}},\ \bibinfo {pages} {018} (\bibinfo {year} {2019})}\BibitemShut {NoStop}%
\bibitem [{\citenamefont {Boyle}\ and\ \citenamefont {Steinhardt}(2022)}]{BoyleSteinhardt22}%
  \BibitemOpen
  \bibfield  {author} {\bibinfo {author} {\bibfnamefont {L.}~\bibnamefont {Boyle}}\ and\ \bibinfo {author} {\bibfnamefont {P.~J.}\ \bibnamefont {Steinhardt}},\ }\href {https://doi.org/10.1103/PhysRevB.106.144113} {\bibfield  {journal} {\bibinfo  {journal} {Phys. Rev. B}\ }\textbf {\bibinfo {volume} {106}},\ \bibinfo {pages} {144113} (\bibinfo {year} {2022})}\BibitemShut {NoStop}%
\bibitem [{\citenamefont {Bellissard}(1992)}]{bellissard_gap_1992}%
  \BibitemOpen
  \bibfield  {author} {\bibinfo {author} {\bibfnamefont {J.}~\bibnamefont {Bellissard}},\ }in\ \href {https://doi.org/10.1007/978-3-662-02838-4_12} {\emph {\bibinfo {booktitle} {From {Number} {Theory} to {Physics}}}},\ \bibinfo {editor} {edited by\ \bibinfo {editor} {\bibfnamefont {M.}~\bibnamefont {Waldschmidt}}, \bibinfo {editor} {\bibfnamefont {P.}~\bibnamefont {Moussa}}, \bibinfo {editor} {\bibfnamefont {J.-M.}\ \bibnamefont {Luck}},\ and\ \bibinfo {editor} {\bibfnamefont {C.}~\bibnamefont {Itzykson}}}\ (\bibinfo  {publisher} {Springer},\ \bibinfo {address} {Berlin, Heidelberg},\ \bibinfo {year} {1992})\ pp.\ \bibinfo {pages} {538--630}\BibitemShut {NoStop}%
\bibitem [{\citenamefont {Singh}\ \emph {et~al.}(2024)\citenamefont {Singh}, \citenamefont {Lloyd},\ and\ \citenamefont {Flicker}}]{Singh24}%
  \BibitemOpen
  \bibfield  {author} {\bibinfo {author} {\bibfnamefont {S.}~\bibnamefont {Singh}}, \bibinfo {author} {\bibfnamefont {J.}~\bibnamefont {Lloyd}},\ and\ \bibinfo {author} {\bibfnamefont {F.}~\bibnamefont {Flicker}},\ }\href@noop {} {\bibfield  {journal} {\bibinfo  {journal} {Physical Review X}\ }\textbf {\bibinfo {volume} {14}},\ \bibinfo {pages} {031005} (\bibinfo {year} {2024})}\BibitemShut {NoStop}%
\bibitem [{Note1()}]{Note1}%
  \BibitemOpen
  \bibinfo {note} {When $\theta =0,\pi $, the periodic Kitaev chain has time reversal symmetry and belongs to class BDI with a $\protect \mathbb {Z}$ topological invariant, suggesting the possibility of 2 or more orthogonal MBS at each end. It has been shown that in order to have more than one unpaired Majorana mode at the end of a single-channel 1D superconductor like that described by Eq.~\protect \eqref {eq:general_H}, modes emerge when adding long-range hoppings~\cite {degottardi2013majorana}, which we do not consider here. Additionally, the realistic models featuring spinful electrons, of which Eq.~\protect \eqref {eq:general_H} is a simplification describing one gapped-out spin species, do not feature true Time reversal Symmetry and do not belong to class BDI~\cite {tewari_topological_2012}. For other values of $\theta $, the isolated chain presented in Eq.~\protect \eqref {eq:general_H} is in class D, with a $\protect \mathbb {Z}_2$ topological invariant corresponding to the presence or absence of MBS,
  and $\theta $ can be absorbed in the definition of the Majorana operators~\cite {kitaev_unpaired_2001}.}\BibitemShut {Stop}%
\bibitem [{\citenamefont {Tewari}\ and\ \citenamefont {Sau}(2012)}]{tewari_topological_2012}%
  \BibitemOpen
  \bibfield  {author} {\bibinfo {author} {\bibfnamefont {S.}~\bibnamefont {Tewari}}\ and\ \bibinfo {author} {\bibfnamefont {J.~D.}\ \bibnamefont {Sau}},\ }\href {https://doi.org/10.1103/PhysRevLett.109.150408} {\bibfield  {journal} {\bibinfo  {journal} {Physical Review Letters}\ }\textbf {\bibinfo {volume} {109}},\ \bibinfo {pages} {150408} (\bibinfo {year} {2012})}\BibitemShut {NoStop}%
\bibitem [{\citenamefont {DeGottardi}\ \emph {et~al.}(2013{\natexlab{a}})\citenamefont {DeGottardi}, \citenamefont {Sen},\ and\ \citenamefont {Vishveshwara}}]{degottardi_majorana_2013}%
  \BibitemOpen
  \bibfield  {author} {\bibinfo {author} {\bibfnamefont {W.}~\bibnamefont {DeGottardi}}, \bibinfo {author} {\bibfnamefont {D.}~\bibnamefont {Sen}},\ and\ \bibinfo {author} {\bibfnamefont {S.}~\bibnamefont {Vishveshwara}},\ }\href {https://doi.org/10.1103/PhysRevLett.110.146404} {\bibfield  {journal} {\bibinfo  {journal} {Physical Review Letters}\ }\textbf {\bibinfo {volume} {110}},\ \bibinfo {pages} {146404} (\bibinfo {year} {2013}{\natexlab{a}})}\BibitemShut {NoStop}%
\bibitem [{\citenamefont {Macé}\ \emph {et~al.}(2016)\citenamefont {Macé}, \citenamefont {Jagannathan},\ and\ \citenamefont {Piéchon}}]{mace_fractal_2016}%
  \BibitemOpen
  \bibfield  {author} {\bibinfo {author} {\bibfnamefont {N.}~\bibnamefont {Macé}}, \bibinfo {author} {\bibfnamefont {A.}~\bibnamefont {Jagannathan}},\ and\ \bibinfo {author} {\bibfnamefont {F.}~\bibnamefont {Piéchon}},\ }\href {https://doi.org/10.1103/PhysRevB.93.205153} {\bibfield  {journal} {\bibinfo  {journal} {Physical Review B}\ }\textbf {\bibinfo {volume} {93}},\ \bibinfo {pages} {205153} (\bibinfo {year} {2016})}\BibitemShut {NoStop}%
\bibitem [{\citenamefont {Jagannathan}(2021)}]{jagannathan_fibonacci_2021}%
  \BibitemOpen
  \bibfield  {author} {\bibinfo {author} {\bibfnamefont {A.}~\bibnamefont {Jagannathan}},\ }\href {https://doi.org/10.1103/RevModPhys.93.045001} {\bibfield  {journal} {\bibinfo  {journal} {Reviews of Modern Physics}\ }\textbf {\bibinfo {volume} {93}},\ \bibinfo {pages} {045001} (\bibinfo {year} {2021})}\BibitemShut {NoStop}%
\bibitem [{\citenamefont {Kraus}\ \emph {et~al.}(2012)\citenamefont {Kraus}, \citenamefont {Lahini}, \citenamefont {Ringel}, \citenamefont {Verbin},\ and\ \citenamefont {Zilberberg}}]{kraus_topological_2012}%
  \BibitemOpen
  \bibfield  {author} {\bibinfo {author} {\bibfnamefont {Y.~E.}\ \bibnamefont {Kraus}}, \bibinfo {author} {\bibfnamefont {Y.}~\bibnamefont {Lahini}}, \bibinfo {author} {\bibfnamefont {Z.}~\bibnamefont {Ringel}}, \bibinfo {author} {\bibfnamefont {M.}~\bibnamefont {Verbin}},\ and\ \bibinfo {author} {\bibfnamefont {O.}~\bibnamefont {Zilberberg}},\ }\href {https://doi.org/10.1103/PhysRevLett.109.106402} {\bibfield  {journal} {\bibinfo  {journal} {Physical Review Letters}\ }\textbf {\bibinfo {volume} {109}},\ \bibinfo {pages} {106402} (\bibinfo {year} {2012})}\BibitemShut {NoStop}%
\bibitem [{\citenamefont {Verbin}\ \emph {et~al.}(2013)\citenamefont {Verbin}, \citenamefont {Zilberberg}, \citenamefont {Kraus}, \citenamefont {Lahini},\ and\ \citenamefont {Silberberg}}]{verbin_observation_2013}%
  \BibitemOpen
  \bibfield  {author} {\bibinfo {author} {\bibfnamefont {M.}~\bibnamefont {Verbin}}, \bibinfo {author} {\bibfnamefont {O.}~\bibnamefont {Zilberberg}}, \bibinfo {author} {\bibfnamefont {Y.~E.}\ \bibnamefont {Kraus}}, \bibinfo {author} {\bibfnamefont {Y.}~\bibnamefont {Lahini}},\ and\ \bibinfo {author} {\bibfnamefont {Y.}~\bibnamefont {Silberberg}},\ }\href {https://doi.org/10.1103/PhysRevLett.110.076403} {\bibfield  {journal} {\bibinfo  {journal} {Physical Review Letters}\ }\textbf {\bibinfo {volume} {110}},\ \bibinfo {pages} {076403} (\bibinfo {year} {2013})}\BibitemShut {NoStop}%
\bibitem [{\citenamefont {Huang}\ and\ \citenamefont {Liu}(2018)}]{huang_quantum_2018}%
  \BibitemOpen
  \bibfield  {author} {\bibinfo {author} {\bibfnamefont {H.}~\bibnamefont {Huang}}\ and\ \bibinfo {author} {\bibfnamefont {F.}~\bibnamefont {Liu}},\ }\href {https://doi.org/10.1103/PhysRevLett.121.126401} {\bibfield  {journal} {\bibinfo  {journal} {Physical Review Letters}\ }\textbf {\bibinfo {volume} {121}},\ \bibinfo {pages} {126401} (\bibinfo {year} {2018})}\BibitemShut {NoStop}%
\bibitem [{\citenamefont {Huang}\ and\ \citenamefont {Liu}(2019)}]{huang_comparison_2019}%
  \BibitemOpen
  \bibfield  {author} {\bibinfo {author} {\bibfnamefont {H.}~\bibnamefont {Huang}}\ and\ \bibinfo {author} {\bibfnamefont {F.}~\bibnamefont {Liu}},\ }\href {https://doi.org/10.1103/PhysRevB.100.085119} {\bibfield  {journal} {\bibinfo  {journal} {Physical Review B}\ }\textbf {\bibinfo {volume} {100}},\ \bibinfo {pages} {085119} (\bibinfo {year} {2019})}\BibitemShut {NoStop}%
\bibitem [{\citenamefont {Macé}\ \emph {et~al.}(2017{\natexlab{b}})\citenamefont {Macé}, \citenamefont {Jagannathan},\ and\ \citenamefont {Piéchon}}]{mace_gap_2017}%
  \BibitemOpen
  \bibfield  {author} {\bibinfo {author} {\bibfnamefont {N.}~\bibnamefont {Macé}}, \bibinfo {author} {\bibfnamefont {A.}~\bibnamefont {Jagannathan}},\ and\ \bibinfo {author} {\bibfnamefont {F.}~\bibnamefont {Piéchon}},\ }\href {https://doi.org/10.1088/1742-6596/809/1/012023} {\bibfield  {journal} {\bibinfo  {journal} {Journal of Physics: Conference Series}\ }\textbf {\bibinfo {volume} {809}},\ \bibinfo {pages} {012023} (\bibinfo {year} {2017}{\natexlab{b}})}\BibitemShut {NoStop}%
\bibitem [{Note2()}]{Note2}%
  \BibitemOpen
  \bibinfo {note} {For continuously varying potentials this invariant is derived from the Chern number in the parent 2D lattice~\cite {flicker_quasiperiodicity_2015}, solving the $q$-sign ambiguity faced when using the Diophantine equation, Eq.~\protect \eqref {eq:energy_gap_labelling}, alone~\cite {avron2014study}. However, for discontinuously varying potentials, as we have in Eq.~\protect \eqref {eq:general_H}, the spectral projection is non-differentiable (see Appendix~\ref {app:phasonWinding}) and so the Berry phase is not well-defined~\cite {band2025quasiperiodicity}, invalidating any reference to the sign of the Berry phase to overcome the $q$-sign ambiguity. For our purposes here, we do not encounter $q$-sign ambiguity at the numerical resolution used, so we proceed with the gap-labelling theorem alone.}\BibitemShut {Stop}%
\bibitem [{\citenamefont {Bagrets}\ and\ \citenamefont {Altland}(2012)}]{bagrets_class_2012}%
  \BibitemOpen
  \bibfield  {author} {\bibinfo {author} {\bibfnamefont {D.}~\bibnamefont {Bagrets}}\ and\ \bibinfo {author} {\bibfnamefont {A.}~\bibnamefont {Altland}},\ }\href {https://doi.org/10.1103/PhysRevLett.109.227005} {\bibfield  {journal} {\bibinfo  {journal} {Physical Review Letters}\ }\textbf {\bibinfo {volume} {109}},\ \bibinfo {pages} {227005} (\bibinfo {year} {2012})}\BibitemShut {NoStop}%
\bibitem [{\citenamefont {Liu}\ \emph {et~al.}(2017)\citenamefont {Liu}, \citenamefont {Sau}, \citenamefont {Stanescu},\ and\ \citenamefont {Das~Sarma}}]{liu_andreev_2017}%
  \BibitemOpen
  \bibfield  {author} {\bibinfo {author} {\bibfnamefont {C.-X.}\ \bibnamefont {Liu}}, \bibinfo {author} {\bibfnamefont {J.~D.}\ \bibnamefont {Sau}}, \bibinfo {author} {\bibfnamefont {T.~D.}\ \bibnamefont {Stanescu}},\ and\ \bibinfo {author} {\bibfnamefont {S.}~\bibnamefont {Das~Sarma}},\ }\href {https://doi.org/10.1103/PhysRevB.96.075161} {\bibfield  {journal} {\bibinfo  {journal} {Physical Review B}\ }\textbf {\bibinfo {volume} {96}},\ \bibinfo {pages} {075161} (\bibinfo {year} {2017})}\BibitemShut {NoStop}%
\bibitem [{\citenamefont {Akhmerov}(2011)}]{akhmerov_quantized_2011}%
  \BibitemOpen
  \bibfield  {author} {\bibinfo {author} {\bibfnamefont {A.~R.}\ \bibnamefont {Akhmerov}},\ }\bibfield  {journal} {\bibinfo  {journal} {Physical Review Letters}\ }\textbf {\bibinfo {volume} {106}},\ \href {https://doi.org/10.1103/PhysRevLett.106.057001} {10.1103/PhysRevLett.106.057001} (\bibinfo {year} {2011})\BibitemShut {NoStop}%
\bibitem [{\citenamefont {Fulga}(2012)}]{fulga_scattering_2012}%
  \BibitemOpen
  \bibfield  {author} {\bibinfo {author} {\bibfnamefont {I.~C.}\ \bibnamefont {Fulga}},\ }\bibfield  {journal} {\bibinfo  {journal} {Physical Review B}\ }\textbf {\bibinfo {volume} {85}},\ \href {https://doi.org/10.1103/PhysRevB.85.165409} {10.1103/PhysRevB.85.165409} (\bibinfo {year} {2012})\BibitemShut {NoStop}%
\bibitem [{\citenamefont {Hegde}\ and\ \citenamefont {Vishveshwara}(2016)}]{hegde_majorana_2016}%
  \BibitemOpen
  \bibfield  {author} {\bibinfo {author} {\bibfnamefont {S.}~\bibnamefont {Hegde}}\ and\ \bibinfo {author} {\bibfnamefont {S.}~\bibnamefont {Vishveshwara}},\ }\href {https://doi.org/10.1103/PhysRevB.94.115166} {\bibfield  {journal} {\bibinfo  {journal} {Physical Review B}\ }\textbf {\bibinfo {volume} {94}},\ \bibinfo {pages} {115166} (\bibinfo {year} {2016})}\BibitemShut {NoStop}%
\bibitem [{\citenamefont {Deng}\ \emph {et~al.}(2016)\citenamefont {Deng}, \citenamefont {Vaitiekėnas}, \citenamefont {Hansen}, \citenamefont {Danon}, \citenamefont {Leijnse}, \citenamefont {Flensberg}, \citenamefont {Nygård}, \citenamefont {Krogstrup},\ and\ \citenamefont {Marcus}}]{deng_majorana_2016}%
  \BibitemOpen
  \bibfield  {author} {\bibinfo {author} {\bibfnamefont {M.~T.}\ \bibnamefont {Deng}}, \bibinfo {author} {\bibfnamefont {S.}~\bibnamefont {Vaitiekėnas}}, \bibinfo {author} {\bibfnamefont {E.~B.}\ \bibnamefont {Hansen}}, \bibinfo {author} {\bibfnamefont {J.}~\bibnamefont {Danon}}, \bibinfo {author} {\bibfnamefont {M.}~\bibnamefont {Leijnse}}, \bibinfo {author} {\bibfnamefont {K.}~\bibnamefont {Flensberg}}, \bibinfo {author} {\bibfnamefont {J.}~\bibnamefont {Nygård}}, \bibinfo {author} {\bibfnamefont {P.}~\bibnamefont {Krogstrup}},\ and\ \bibinfo {author} {\bibfnamefont {C.~M.}\ \bibnamefont {Marcus}},\ }\href {https://doi.org/10.1126/science.aaf3961} {\bibfield  {journal} {\bibinfo  {journal} {Science}\ }\textbf {\bibinfo {volume} {354}},\ \bibinfo {pages} {1557} (\bibinfo {year} {2016})}\BibitemShut {NoStop}%
\bibitem [{\citenamefont {Vuik}\ \emph {et~al.}(2019)\citenamefont {Vuik}, \citenamefont {Nijholt}, \citenamefont {Akhmerov},\ and\ \citenamefont {Wimmer}}]{vuik_reproducing_2019}%
  \BibitemOpen
  \bibfield  {author} {\bibinfo {author} {\bibfnamefont {A.}~\bibnamefont {Vuik}}, \bibinfo {author} {\bibfnamefont {B.}~\bibnamefont {Nijholt}}, \bibinfo {author} {\bibfnamefont {A.~R.}\ \bibnamefont {Akhmerov}},\ and\ \bibinfo {author} {\bibfnamefont {M.}~\bibnamefont {Wimmer}},\ }\href {https://doi.org/10.21468/SciPostPhys.7.5.061} {\bibfield  {journal} {\bibinfo  {journal} {SciPost Physics}\ }\textbf {\bibinfo {volume} {7}},\ \bibinfo {pages} {061} (\bibinfo {year} {2019})}\BibitemShut {NoStop}%
\bibitem [{\citenamefont {Pan}\ and\ \citenamefont {Das~Sarma}(2020)}]{pan_physical_2020}%
  \BibitemOpen
  \bibfield  {author} {\bibinfo {author} {\bibfnamefont {H.}~\bibnamefont {Pan}}\ and\ \bibinfo {author} {\bibfnamefont {S.}~\bibnamefont {Das~Sarma}},\ }\href {https://doi.org/10.1103/PhysRevResearch.2.013377} {\bibfield  {journal} {\bibinfo  {journal} {Physical Review Research}\ }\textbf {\bibinfo {volume} {2}},\ \bibinfo {pages} {013377} (\bibinfo {year} {2020})}\BibitemShut {NoStop}%
\bibitem [{\citenamefont {Sticlet}\ \emph {et~al.}(2012)\citenamefont {Sticlet}, \citenamefont {Bena},\ and\ \citenamefont {Simon}}]{sticlet_spin_2012}%
  \BibitemOpen
  \bibfield  {author} {\bibinfo {author} {\bibfnamefont {D.}~\bibnamefont {Sticlet}}, \bibinfo {author} {\bibfnamefont {C.}~\bibnamefont {Bena}},\ and\ \bibinfo {author} {\bibfnamefont {P.}~\bibnamefont {Simon}},\ }\href {https://doi.org/10.1103/PhysRevLett.108.096802} {\bibfield  {journal} {\bibinfo  {journal} {Physical Review Letters}\ }\textbf {\bibinfo {volume} {108}},\ \bibinfo {pages} {096802} (\bibinfo {year} {2012})}\BibitemShut {NoStop}%
\bibitem [{\citenamefont {Sedlmayr}\ and\ \citenamefont {Bena}(2015)}]{sedlmayr_visualizing_2015}%
  \BibitemOpen
  \bibfield  {author} {\bibinfo {author} {\bibfnamefont {N.}~\bibnamefont {Sedlmayr}}\ and\ \bibinfo {author} {\bibfnamefont {C.}~\bibnamefont {Bena}},\ }\href {https://doi.org/10.1103/PhysRevB.92.115115} {\bibfield  {journal} {\bibinfo  {journal} {Physical Review B}\ }\textbf {\bibinfo {volume} {92}},\ \bibinfo {pages} {115115} (\bibinfo {year} {2015})}\BibitemShut {NoStop}%
\bibitem [{\citenamefont {Awoga}\ and\ \citenamefont {Cayao}(2024)}]{awoga_identifying_2024}%
  \BibitemOpen
  \bibfield  {author} {\bibinfo {author} {\bibfnamefont {O.~A.}\ \bibnamefont {Awoga}}\ and\ \bibinfo {author} {\bibfnamefont {J.}~\bibnamefont {Cayao}},\ }\href {https://doi.org/10.1103/PhysRevB.110.165404} {\bibfield  {journal} {\bibinfo  {journal} {Physical Review B}\ }\textbf {\bibinfo {volume} {110}},\ \bibinfo {pages} {165404} (\bibinfo {year} {2024})}\BibitemShut {NoStop}%
\bibitem [{\citenamefont {Damanik}\ and\ \citenamefont {Lenz}(1999)}]{damanik_uniform_1999}%
  \BibitemOpen
  \bibfield  {author} {\bibinfo {author} {\bibfnamefont {D.}~\bibnamefont {Damanik}}\ and\ \bibinfo {author} {\bibfnamefont {D.}~\bibnamefont {Lenz}},\ }\href {https://doi.org/10.1007/s002200050742} {\bibfield  {journal} {\bibinfo  {journal} {Communications in Mathematical Physics}\ }\textbf {\bibinfo {volume} {207}},\ \bibinfo {pages} {687} (\bibinfo {year} {1999})}\BibitemShut {NoStop}%
\bibitem [{\citenamefont {Pytheas~Fogg}(2002)}]{NPytheasFogg}%
  \BibitemOpen
  \bibfield  {author} {\bibinfo {author} {\bibfnamefont {N.}~\bibnamefont {Pytheas~Fogg}},\ }\href@noop {} {\emph {\bibinfo {title} {Substitutions in dynamics, arithmetics and combinatorics}}},\ edited by\ \bibinfo {editor} {\bibfnamefont {V.}~\bibnamefont {Berthé}}, \bibinfo {editor} {\bibfnamefont {S.}~\bibnamefont {Ferenczi}}, \bibinfo {editor} {\bibfnamefont {C.}~\bibnamefont {Mauduit}},\ and\ \bibinfo {editor} {\bibfnamefont {A.}~\bibnamefont {Siegel}},\ \bibinfo {series} {Lecture Notes in Mathematics}, Vol.\ \bibinfo {volume} {1794}\ (\bibinfo  {publisher} {Springer-Verlag (Berlin)},\ \bibinfo {year} {2002})\BibitemShut {NoStop}%
\bibitem [{\citenamefont {DeGottardi}\ \emph {et~al.}(2013{\natexlab{b}})\citenamefont {DeGottardi}, \citenamefont {Thakurathi}, \citenamefont {Vishveshwara},\ and\ \citenamefont {Sen}}]{degottardi2013majorana}%
  \BibitemOpen
  \bibfield  {author} {\bibinfo {author} {\bibfnamefont {W.}~\bibnamefont {DeGottardi}}, \bibinfo {author} {\bibfnamefont {M.}~\bibnamefont {Thakurathi}}, \bibinfo {author} {\bibfnamefont {S.}~\bibnamefont {Vishveshwara}},\ and\ \bibinfo {author} {\bibfnamefont {D.}~\bibnamefont {Sen}},\ }\href@noop {} {\bibfield  {journal} {\bibinfo  {journal} {Physical Review B—Condensed Matter and Materials Physics}\ }\textbf {\bibinfo {volume} {88}},\ \bibinfo {pages} {165111} (\bibinfo {year} {2013}{\natexlab{b}})}\BibitemShut {NoStop}%
\bibitem [{\citenamefont {Hofstadter}(1976)}]{hofstadter_energy_1976}%
  \BibitemOpen
  \bibfield  {author} {\bibinfo {author} {\bibfnamefont {D.~R.}\ \bibnamefont {Hofstadter}},\ }\href {https://doi.org/10.1103/PhysRevB.14.2239} {\bibfield  {journal} {\bibinfo  {journal} {Physical Review B}\ }\textbf {\bibinfo {volume} {14}},\ \bibinfo {pages} {2239} (\bibinfo {year} {1976})}\BibitemShut {NoStop}%
\bibitem [{\citenamefont {Simon}(1982)}]{simon_almost_1982}%
  \BibitemOpen
  \bibfield  {author} {\bibinfo {author} {\bibfnamefont {B.}~\bibnamefont {Simon}},\ }\href {https://doi.org/10.1016/S0196-8858(82)80018-3} {\bibfield  {journal} {\bibinfo  {journal} {Advances in Applied Mathematics}\ }\textbf {\bibinfo {volume} {3}},\ \bibinfo {pages} {463} (\bibinfo {year} {1982})}\BibitemShut {NoStop}%
\bibitem [{\citenamefont {Thouless}\ \emph {et~al.}(1982)\citenamefont {Thouless}, \citenamefont {Kohmoto}, \citenamefont {Nightingale},\ and\ \citenamefont {Den~Nijs}}]{thouless_quantized_1982}%
  \BibitemOpen
  \bibfield  {author} {\bibinfo {author} {\bibfnamefont {D.~J.}\ \bibnamefont {Thouless}}, \bibinfo {author} {\bibfnamefont {M.}~\bibnamefont {Kohmoto}}, \bibinfo {author} {\bibfnamefont {M.~P.}\ \bibnamefont {Nightingale}},\ and\ \bibinfo {author} {\bibfnamefont {M.}~\bibnamefont {Den~Nijs}},\ }\href {https://doi.org/10.1103/PhysRevLett.49.405} {\bibfield  {journal} {\bibinfo  {journal} {Physical Review Letters}\ }\textbf {\bibinfo {volume} {49}},\ \bibinfo {pages} {405} (\bibinfo {year} {1982})}\BibitemShut {NoStop}%
\bibitem [{\citenamefont {Kohmoto}(1985)}]{kohmoto_topological_1985}%
  \BibitemOpen
  \bibfield  {author} {\bibinfo {author} {\bibfnamefont {M.}~\bibnamefont {Kohmoto}},\ }\href {https://doi.org/10.1016/0003-4916(85)90148-4} {\bibfield  {journal} {\bibinfo  {journal} {Annals of Physics}\ }\textbf {\bibinfo {volume} {160}},\ \bibinfo {pages} {343} (\bibinfo {year} {1985})}\BibitemShut {NoStop}%
\bibitem [{\citenamefont {Dana}\ \emph {et~al.}(1985)\citenamefont {Dana}, \citenamefont {Avron},\ and\ \citenamefont {Zak}}]{dana_quantised_1985}%
  \BibitemOpen
  \bibfield  {author} {\bibinfo {author} {\bibfnamefont {I.}~\bibnamefont {Dana}}, \bibinfo {author} {\bibfnamefont {Y.}~\bibnamefont {Avron}},\ and\ \bibinfo {author} {\bibfnamefont {J.}~\bibnamefont {Zak}},\ }\href {https://doi.org/10.1088/0022-3719/18/22/004} {\bibfield  {journal} {\bibinfo  {journal} {Journal of Physics C: Solid State Physics}\ }\textbf {\bibinfo {volume} {18}},\ \bibinfo {pages} {L679} (\bibinfo {year} {1985})}\BibitemShut {NoStop}%
\bibitem [{\citenamefont {Bellissard}(1986)}]{bellissard_k-theory_1986}%
  \BibitemOpen
  \bibfield  {author} {\bibinfo {author} {\bibfnamefont {J.}~\bibnamefont {Bellissard}},\ }in\ \href {https://doi.org/10.1007/3-540-16777-3_74} {\emph {\bibinfo {booktitle} {Statistical {Mechanics} and {Field} {Theory}: {Mathematical} {Aspects}}}},\ \bibinfo {editor} {edited by\ \bibinfo {editor} {\bibfnamefont {T.~C.}\ \bibnamefont {Dorlas}}, \bibinfo {editor} {\bibfnamefont {N.~M.}\ \bibnamefont {Hugenholtz}},\ and\ \bibinfo {editor} {\bibfnamefont {M.}~\bibnamefont {Winnink}}}\ (\bibinfo  {publisher} {Springer},\ \bibinfo {address} {Berlin, Heidelberg},\ \bibinfo {year} {1986})\ pp.\ \bibinfo {pages} {99--156}\BibitemShut {NoStop}%
\bibitem [{\citenamefont {Abanov}\ \emph {et~al.}(1998)\citenamefont {Abanov}, \citenamefont {Talstra},\ and\ \citenamefont {Wiegmann}}]{abanov1998hierarchical}%
  \BibitemOpen
  \bibfield  {author} {\bibinfo {author} {\bibfnamefont {A.}~\bibnamefont {Abanov}}, \bibinfo {author} {\bibfnamefont {J.}~\bibnamefont {Talstra}},\ and\ \bibinfo {author} {\bibfnamefont {P.}~\bibnamefont {Wiegmann}},\ }\href@noop {} {\bibfield  {journal} {\bibinfo  {journal} {Nuclear Physics B}\ }\textbf {\bibinfo {volume} {525}},\ \bibinfo {pages} {571} (\bibinfo {year} {1998})}\BibitemShut {NoStop}%
\bibitem [{\citenamefont {Satija}(2016)}]{satija2016tale}%
  \BibitemOpen
  \bibfield  {author} {\bibinfo {author} {\bibfnamefont {I.~I.}\ \bibnamefont {Satija}},\ }\href@noop {} {\bibfield  {journal} {\bibinfo  {journal} {The European Physical Journal Special Topics}\ }\textbf {\bibinfo {volume} {225}},\ \bibinfo {pages} {2533} (\bibinfo {year} {2016})}\BibitemShut {NoStop}%
\bibitem [{\citenamefont {Beckus}\ \emph {et~al.}(2025)\citenamefont {Beckus}, \citenamefont {Bellissard},\ and\ \citenamefont {Thomas}}]{beckus2025spectral}%
  \BibitemOpen
  \bibfield  {author} {\bibinfo {author} {\bibfnamefont {S.}~\bibnamefont {Beckus}}, \bibinfo {author} {\bibfnamefont {J.}~\bibnamefont {Bellissard}},\ and\ \bibinfo {author} {\bibfnamefont {Y.}~\bibnamefont {Thomas}},\ }in\ \href@noop {} {\emph {\bibinfo {booktitle} {Annales Henri Poincar{\'e}}}}\ (\bibinfo {organization} {Springer},\ \bibinfo {year} {2025})\ pp.\ \bibinfo {pages} {1--38}\BibitemShut {NoStop}%
\bibitem [{\citenamefont {Jitomirskaya}\ \emph {et~al.}(2022)\citenamefont {Jitomirskaya}, \citenamefont {Konstantinov}, \citenamefont {Krasovsky} \emph {et~al.}}]{jitomirskaya2022spectrum}%
  \BibitemOpen
  \bibfield  {author} {\bibinfo {author} {\bibfnamefont {S.}~\bibnamefont {Jitomirskaya}}, \bibinfo {author} {\bibfnamefont {L.}~\bibnamefont {Konstantinov}}, \bibinfo {author} {\bibfnamefont {I.}~\bibnamefont {Krasovsky}}, \emph {et~al.},\ }\href@noop {} {\bibfield  {journal} {\bibinfo  {journal} {Journal of Spectral Theory}\ }\textbf {\bibinfo {volume} {12}},\ \bibinfo {pages} {11} (\bibinfo {year} {2022})}\BibitemShut {NoStop}%
\bibitem [{\citenamefont {Kohmoto}\ \emph {et~al.}(1983)\citenamefont {Kohmoto}, \citenamefont {Kadanoff},\ and\ \citenamefont {Tang}}]{kohmoto1983localization}%
  \BibitemOpen
  \bibfield  {author} {\bibinfo {author} {\bibfnamefont {M.}~\bibnamefont {Kohmoto}}, \bibinfo {author} {\bibfnamefont {L.~P.}\ \bibnamefont {Kadanoff}},\ and\ \bibinfo {author} {\bibfnamefont {C.}~\bibnamefont {Tang}},\ }\href@noop {} {\bibfield  {journal} {\bibinfo  {journal} {Physical Review Letters}\ }\textbf {\bibinfo {volume} {50}},\ \bibinfo {pages} {1870} (\bibinfo {year} {1983})}\BibitemShut {NoStop}%
\bibitem [{\citenamefont {Band}\ and\ \citenamefont {Beckus}(2025)}]{band2025quasiperiodicity}%
  \BibitemOpen
  \bibfield  {author} {\bibinfo {author} {\bibfnamefont {R.}~\bibnamefont {Band}}\ and\ \bibinfo {author} {\bibfnamefont {S.}~\bibnamefont {Beckus}},\ }\href@noop {} {\bibfield  {journal} {\bibinfo  {journal} {arXiv preprint arXiv:2509.24025}\ } (\bibinfo {year} {2025})}\BibitemShut {NoStop}%
\bibitem [{\citenamefont {Madsen}\ \emph {et~al.}(2013)\citenamefont {Madsen}, \citenamefont {Bergholtz},\ and\ \citenamefont {Brouwer}}]{Madsen13}%
  \BibitemOpen
  \bibfield  {author} {\bibinfo {author} {\bibfnamefont {K.~A.}\ \bibnamefont {Madsen}}, \bibinfo {author} {\bibfnamefont {E.~J.}\ \bibnamefont {Bergholtz}},\ and\ \bibinfo {author} {\bibfnamefont {P.~W.}\ \bibnamefont {Brouwer}},\ }\href {https://doi.org/10.1103/PhysRevB.88.125118} {\bibfield  {journal} {\bibinfo  {journal} {Phys. Rev. B}\ }\textbf {\bibinfo {volume} {88}},\ \bibinfo {pages} {125118} (\bibinfo {year} {2013})}\BibitemShut {NoStop}%
\bibitem [{\citenamefont {Avron}\ \emph {et~al.}(2014)\citenamefont {Avron}, \citenamefont {Kenneth},\ and\ \citenamefont {Yehoshua}}]{avron2014study}%
  \BibitemOpen
  \bibfield  {author} {\bibinfo {author} {\bibfnamefont {J.}~\bibnamefont {Avron}}, \bibinfo {author} {\bibfnamefont {O.}~\bibnamefont {Kenneth}},\ and\ \bibinfo {author} {\bibfnamefont {G.}~\bibnamefont {Yehoshua}},\ }\href@noop {} {\bibfield  {journal} {\bibinfo  {journal} {Journal of Physics A: Mathematical and Theoretical}\ }\textbf {\bibinfo {volume} {47}},\ \bibinfo {pages} {185202} (\bibinfo {year} {2014})}\BibitemShut {NoStop}%
\end{thebibliography}

\appendix

\counterwithin{figure}{section}

\section{\label{app:phasonWinding}Phason Winding and Topologically Trivial mid-gap States}
\begin{figure}
    \centering
    \includegraphics[width=\linewidth]{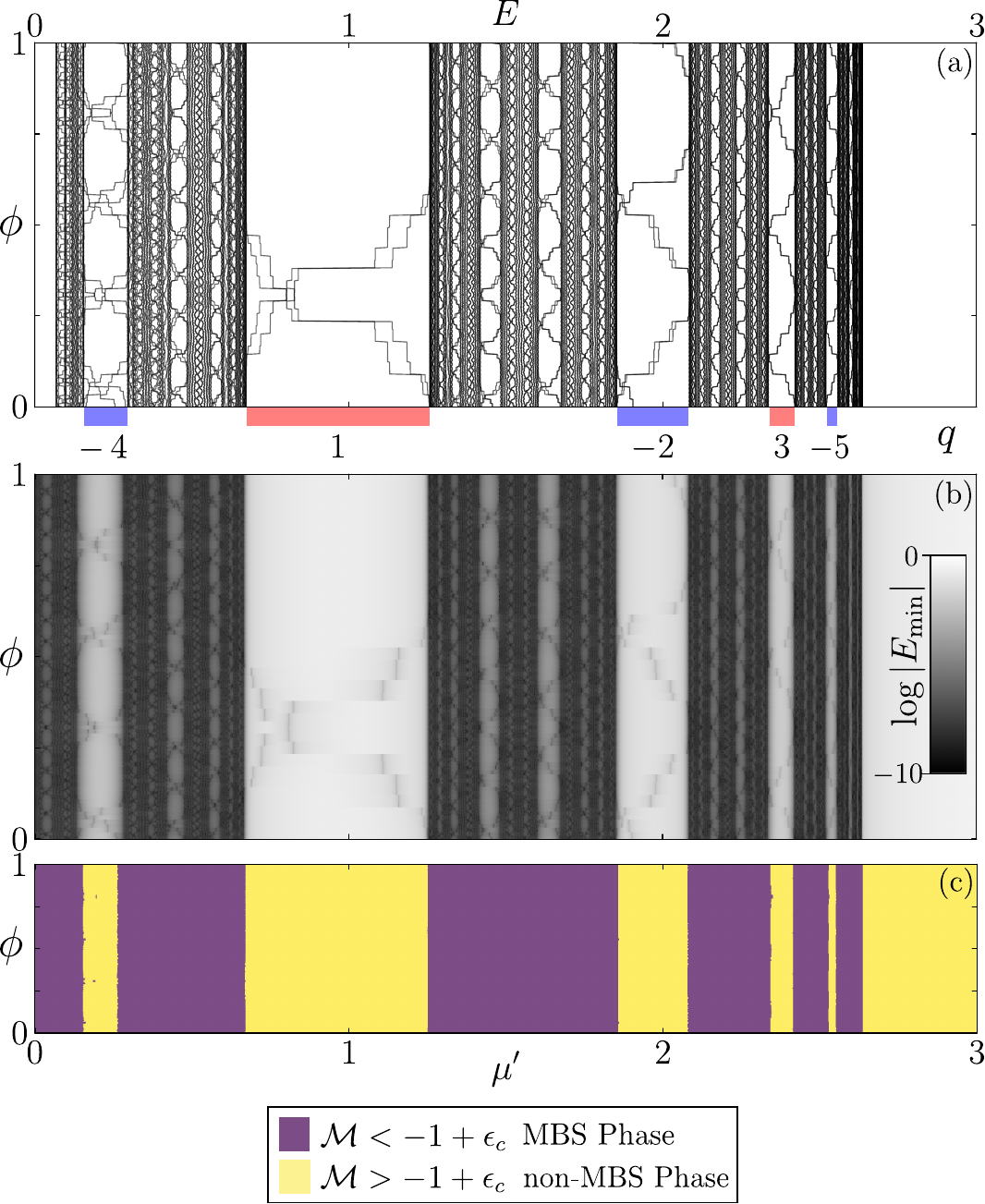}
    \caption{Energy spectrum \textbf{(a)} plotted for fixed $\gamma = \varphi^{-1}$ over a full phason period $0<\phi\leq1$ showing the how individual mid-gap states exhibit integer winding labelled by $q$, taking positive (red) or negative (blue) values. \textbf{(b)} shows the minimum positive energy $(|E_\text{min}|)$ as a function of $\mu'$ and $\phi$, revealing the winding behaviour of the trivial zero-energy states induced by their energy mid-gap-state counterparts shown in the top panel. The MBS topology is undisturbed by $\phi$ for $\epsilon_c=10^{-3}$ \textbf{(c)}.}
    \label{fig:phasonWinding}
\end{figure}

Within the fractal energy structure of a quasicrystalline spectrum there are isolated mid-gap states which have distinctly different behaviour to the main spectrum when the phason angle is varied. As $\phi$ is varied these mid-gap states wind across energy gaps, producing a quantised winding number over a full period $0\leq\phi<1$. Meanwhile, the rest of the energy spectrum that defines the background gap structure is fixed by $\gamma$, remaining constant under variations of $\phi$. This can be seen in Fig.~\ref{fig:phasonWinding}~(a), which also identifies the winding number, $q$, of the gaps satisfying the QC-SC competition criterion in Eq.~\eqref{eq:QC-SCCompCriterion}.

Fig.~\ref{fig:phasonWinding}~(b) shows the smallest positive eigenvalue of $H$ (Eq.~\eqref{eq:general_H}) as a function of $\phi$ and $\mu'$ for $\gamma=\varphi^{-1}$. We observe near-zero-energy states within topologically trivial regions of $\mu'$ that wind around the MBS phase gaps in the same way as their corresponding energy mid-gap states. In these regions, the QC-SC competition criterion in Eq.~\eqref{eq:QC-SCCompCriterion} is satisfied, breaking the topological phase and leading to a trivial Majorana polarisation $\mathcal{M} > -1 + \epsilon_c$. Here, the value of $\mathcal{M}$ is unperturbed by the winding mid-gap states, as can be seen in Fig.~\ref{fig:phasonWinding}~(c) for $\epsilon_c=10^{-3}$. This is a clear indication that these winding states are not Majorana modes, but rather topologically trivial zero-energy states.

In the $\mu'$ regions where the topological phase is not destroyed by the QC gap, the winding mid-gap states lead to a finite hybridisation of the Majorana modes, manifesting as small deviations of the Majorana states from zero energy.

\section{\label{app:hierarchy}Hierarchy of QC-Induced Hybridisation of MBS}

\begin{figure}
    \centering
    \includegraphics[width=\linewidth]{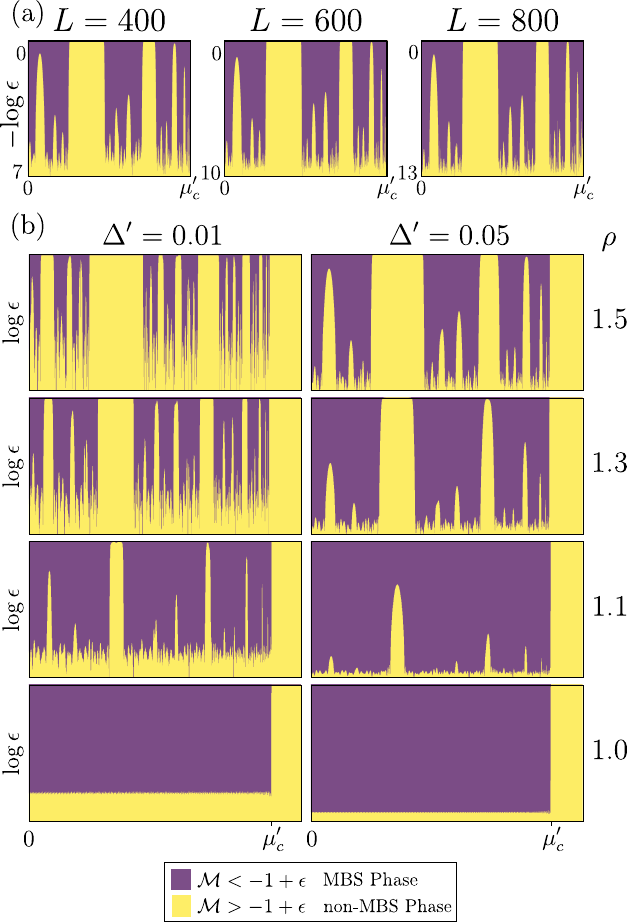}
    \caption{\textbf{(a)} The relative height of the MP domes remains constant whilst varying $L$ for a fixed ratio between QC and SC strength; shown for $\rho=1.5$ and $\Delta'=0.05$ for $L=400$ (left), 600 (middle), and 800 (right). Meanwhile, \textbf{(b)} shows how $\rho$ and $\Delta'$ both change the relative height of domes. The $\log \epsilon$ scales are fixed in each column to illustrate the emergence of new domes and MBS phase gaps as $\rho'/\Delta'$ is increased.}
    \label{fig:hierarchyCombined}
\end{figure}

Here we elucidate the effects of the length $L$, quasicrystal hopping ratio $\rho$, and superconducting pairing $\Delta'$ on the dome-height hierarchy observed in the Majorana polarisation in Fig.~\ref{fig:Figure 2}, ultimately showing that it is the competition between quasicrystal and superconducting strengths which dictates it. This adds clarity to our argument that Majorana's butterfly (MB) is a subset of the fractal structure in Kitaev's butterfly (KB) controlled by $\rho/\Delta'$ and not $L$.

We observe that the relative height of domes of trivial $\mathcal{M}>-1 + \epsilon$ shown in Fig.~\ref{fig:hierarchyCombined}~(a) is unchanged as $L$ is increased. Comparing the hierarchy of Majorana polarisation domes for different values of $L$, we observe they have the same relative height regardless of the length of the chain. Most notably, varying $L$ does not change the number of MBS phase transitions. Regions of $\mu'$ for which the quasicrystallinity overpowers superconductivity, satisfying the QC-SC competition criterion, remain trivial for larger $L$. Therefore, despite the fact that $L$ is the finite-scale control of the fractality of the QC energy spectrum, and hence of KB, it does not have the same control effect on the fractality of the MBS phase transition, leading to MB, where the fractality is controlled by $\rho/\Delta'$.\\

Unlike the effect of $L$ in Fig.~\ref{fig:hierarchyCombined}~(a), the effects of $\rho$ and $\Delta'$ in Fig.~\ref{fig:hierarchyCombined}~(b) reveal a clear change in relative dome heights controlled by $\rho/\Delta'$. Firstly, for $\rho=1.0$ there is no dome structure for any $\Delta'$ as this corresponds to the periodic crystal case $t_0=t_1$; only the single phase transition at $\mu'_c$ is observed. Increasing $\rho$ introduces quasicrystallinity as $t_0 \neq t_1$, and distinct domes emerge. Already, at $\rho=1.1$  for $\Delta'=0.01$ domes have reached up to $\log\epsilon=0$, defining clear breaks in the MBS topological phase, and elsewhere a hierarchy of finite hybridisation of MBS. Increasing $\rho$ further raises the height of those domes at $\log\epsilon<0$, leading to new MBS phase gaps, and introducing new finite-hybridisation domes within the MBS phase. Whilst varying $\rho$ or $\Delta'$ has a different effect on the hierarchy due to the third energy scale of $\mu'$ in the system, it is clear the ratio $\rho/\Delta'$ changes the dome hierarchy and thus the regions which satisfy the QC-SC competition criterion. For MB this means taking a larger subset of the full fractal structure of KB.

\section{\label{app:unnormalisedButterflies}Unnormalised Butterflies}

\begin{figure*}
    \centering
    \includegraphics[width=\textwidth]{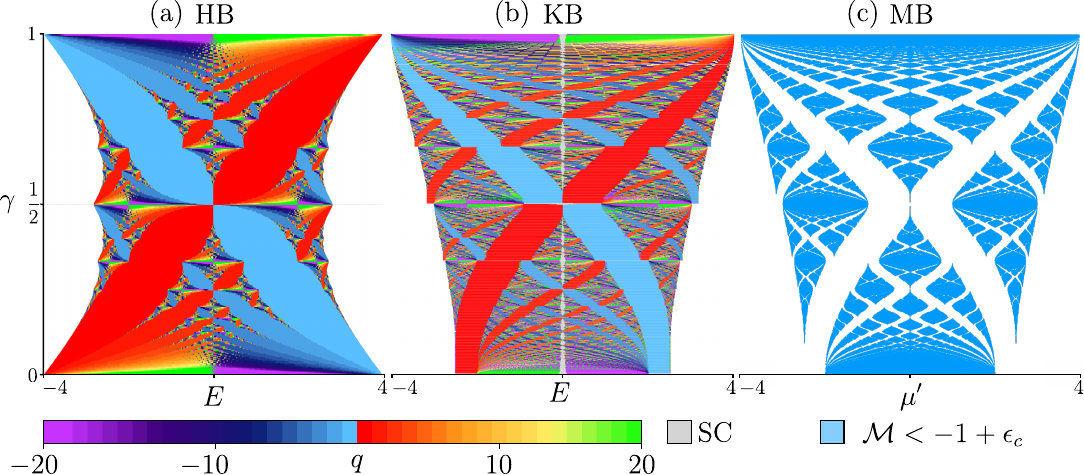}
    \caption{\textbf{(a)} Hofstadter's butterfly (HB) and \textbf{(b)} Kitaev's butterfly (KB) coloured by $q$ and \textbf{(c)} Majorana's butterfly for $L=500$, $\rho=2.0$ and $\Delta'=0.05$. The three structures are presented here unnormalised to emphasise the structural differences that were not relevant to our discussion of topology in Sec.~\ref{sec:Generalising}.}
    \label{fig:A4_unNormalisedButterflies}
\end{figure*}

In Fig.~\ref{fig:Figure 3}, Hofstadter's, Kitaev's and Majorana's butterflies were presented normalised to their bandwidth $W_\gamma$ to emphasise the similarities in topological structure. However, there are further structural differences that do not affect our topological comparison, which can be seen in here in Fig.~\ref{fig:A4_unNormalisedButterflies}. The most notable difference is the symmetry about $\gamma=1/2$ --- HB is symmetric, whereas KB and MB are both asymmetric. This is due to the asymmetry in bandwidth $W_\gamma$ derived from the dependence of $\bar{t}$ on $\gamma$ enforced when $t_0\neq t_1$. The second important difference is the discontinuity of $W_\gamma$ as a function of $\gamma$ in KB. This discontinuity is the result of discontinuously varying potentials as shown by Kohmoto et al.~\cite{kohmoto1983localization} and recently visualised by Band et al.~\cite{band2025quasiperiodicity}.

\end{document}